\documentclass[%
 reprint,
 amsmath,amssymb,
prl,
]{revtex4-1}
\usepackage{natbib}
\usepackage{graphicx}
\usepackage{dcolumn}
\usepackage{bm}
\usepackage{hyperref}
\usepackage[mathlines]{lineno}
\usepackage{verbatim} 
\usepackage{amsmath}
\usepackage{multirow} 
\usepackage{boxhandler}
\usepackage{tabularx}
\usepackage{amsmath}
\usepackage{threeparttable}
\usepackage{gensymb}
\usepackage{textcomp}
\usepackage{soul}
\usepackage{color}

\newcommand{\GCO}{\mbox{GdCrO$_3$~}}

\begin{document}

\preprint{APS/123-QED}

\title{Role of local short-scale correlations in the mechanism of negative magnetization }
\author{Malvika Tripathi$^1$}\author{T. Chatterji$^2$}\author{H. E. Fischer$^2$}\author{R. Raghunathan$^1$}\author{Supriyo Majumder$^1$}\author{R. J. Choudhary$^1$}\author{D. M. Phase$^1$}
\affiliation{ $^1$UGC-DAE Consortium for Scientific Research, Indore-452001, India}
\affiliation{ $^2$Institut Laue-Langevin, 38042 Grenoble Cedex, France}

\begin{abstract}

We elaborate here why the antiferromagnetically ordered GdCrO$_3$ responds in a diamagnetic way under certain conditions, by monitoring the evolution of the microscopic global and local magnetic phases. Using high energy  $\sim$ 0.3 eV neutrons, the magnetic ordering is shown to adopt three distinct magnetic phases at different temperatures: G$_x^{Cr}$,A$_y^{Cr}$,F$_z^{Cr}$ below N\'eel temperature = 171 K; (F$_x^{Cr}$, C$_y^{Cr}$, G$_z^{Cr}$)$\bullet$( F$_x$$^{Gd}$,C$_y$$^{Gd}$) below 7 K and an intermediate phase for 7 K $ \le T \le$ 20 K in the vicinity of spin-reorientation phase transition.  Although, bulk magnetometry reveals a huge negative magnetization (NM) in the terms of both magnitude and temperature range ( $M_{- max}$ ( 18 K)$\sim$ 35 $\times M_{+ max}$ (161 K), $\Delta T \sim 110$ K in presence of $\mu_0H$ = 0.01 T); the long-range magnetic structure and derived ordered moments are unable to explain the NM. Real-space analysis of the total (Bragg's + diffuse) scattering reveals significant magnetic correlations extending up to $\sim$ 9 $\AA$. Accounting for these short-range correlations with a spin model reveals spin frustration in the S= 3 ground state, comprising competing first, second and third next nearest exchange interactions with values J$_1$ = 2.3 K,  J$_2$ = -1.66 K and J$_3$ = 2.19 K in presence of internal field, governs the observance of NM in GdCrO$_{3}$.

\begin{description}

\item[PACS numbers: ]
 75.25.-j, 75.40.-s, 75.50.Ee
\end{description}
\end{abstract}
\maketitle

 Negative magnetization (NM) in the magnetically ordered systems endowing a net magnetization opposite to the applied field besides being a fascinating subject from a fundamental scientific point of view, has also been associated with a number of debates regarding the origin and reproducibility of this phenomena \cite{gu2014physical, ren1998temperature,belik2013fresh}. Since the hypothetical prediction of this phenomenon by N\'eel \cite{neel1948magnetic}, a wide range of observations of NM have been noted in a variety of systems including ferrites, rare-earth garnets, intermetallic alloys, spin chain and layered compounds \cite{menyuk1960magnetization, hong2004spin, hase2011negative,chavan1996magnetization, buschow1971intermetallic}. Depending on the class of materials, the origin of the NM is also diverse; possible reasons include: compensation of the magnetic moments at non-identical magnetic sites in ferrimagnets under the framework of molecular field theory \cite{neel1948magnetic,chikazumi2009physics,smart1955neel}, the imbalance of spin and orbital moments \cite{adachi1999ferromagnet, adachi2001zero}, competition of single ion anisotropy with Dzyaloshinskii-Moriya coupling \cite{ren1998temperature, ren2000magnetic, nguyen1995magnetic} and phase inhomogeneity caused by a very small amount of defects \cite{tung2007magnetization, tung2006tunable}. The perovskite $RBO_3$ ($R$= rare-earth, $B$= transition metal) family, where the difference in magnetic ordering temperatures of $R^{3+}$ and $B^{3+}$ ionic sites is huge ($ ^{B}T_N-^{R}T_N \ge 100$ K), also represents an intriguing class of such materials. Several members of rare-earth orthochromite $R$CrO$_3$, orthoferrite $R$FeO$_3$ and orthovanadate $R$VO$_3$ families are known to realize this situation, either in un-doped form or in chemically substituted compositions \cite{cao2014magnetization, sharannia2017observation, yuan2013spin, zhao2016origin, ren1998temperature}. As the $R^{3+}$ ions are paramagnetic in observed negative magnetization (NM) regime, this class does not directly belong to the N\'eel's oppositely coupled ferrimagnetic materials. In this case, the origin of NM is phenomenologically described by the assumption that the paramagnetic $R^{3+}$ site is polarized by the internal magnetic field (H$_e$) imposed by the magnetically ordered $B^{3+}$ ions and, the two nonequivalent magnetic species $R^{3+}$ and $B^{3+}$ (Cr, Fe, V) are coupled anti-ferromgnetically \cite{su2010magnetization,yoshii2011positive,mao2011temperature,yoshii2000reversal,yoshii2001magnetic}. In the present work, we aim to revisit this hypothesis and origin of NM in distorted orthorhombic perovskite GdCrO$_3$. Reasons justifying this choice include a huge magnitude of observed NM  and  broad temperature span $\Delta T$ $\sim$ 110 K in the presence of applied magnetic field $\mu_0 H$ = 0.01 T, whereas the maximum absolute NM is $\sim$ 35 times larger than the maximum positive moment obtained, providing an ideal scenario for switching equipment. In addition, the observation of NM in GdCrO$_{3}$ also reveals the interesting dependency on choice of measuring route, manifesting different behaviors in cooling and warming cycles. The phenomenological assumption comprising of opposite alignment of polarized para-magnetic Gd$^{3+}$  ions with respect to Cr$^{3+}$ ions is insufficient to explain the observed measuring path dependency of the NM. 
 
 The key reason for the discrepancy is the lack of understanding of the microscopic magnetic structure and its transformation with respect to the temperature. The presence of very high neutron absorbing natural Gd element has been the reason so far to disregard the neutron diffraction measurements. High neutron absorption is a consequence of nuclear resonances of two Gd isotopes present in natural Gd: $^{155}$Gd and $^{157}$Gd at very low energies $E$ = 0.0281 eV and $ E$ = 0.0312 eV respectively, whereas resonance energy width being $\Delta E = 0.105$ eV \cite{lynn1990resonance}. To overcome the high absorption we recorded the neutron diffraction profiles with incident neutron energy tuned to $E = 0.328$ eV ( $\lambda = 0.4994 ~ \AA$ ), a value much higher than the resonance energy width. In this present study, we aim to construct the temperature driven microscopic phase diagram and qualitatively understand the mechanism of NM in GdCrO$_3$. 

Crystal structure, phase purity and valence states of chromium ions are confirmed using X-Ray Diffraction (Bruker D2 PHASER Desktop Diffractometer, Cu-K$\alpha$, $\lambda = 1.54 ~\AA$) and X-ray photoemission spectroscopy (XPS) with Al-K$\alpha$ ($E = 1486.7 $eV) lab source. Details of the sample preparation, crystallographic phase refinement with respect to the Rietveld generated model pattern and valence state confirmation by XPS analysis are discussed in supplementary material (SM)\cite{supplementarymaterial} with the help of references therein\cite{allen1976multiplet, ikemoto1976x}. Magnetometric measurements are performed using commercial SQUID-VSM (MPMS-7 T, Quantum Design, USA). Temperature dependent magnetization $\textit{M}$(T) is measured in the conventional zero field cooled (ZFC), field cooled cooling (FCC) and field cooled warming (FCW) protocols. Before each $\textit{M}$(T) measurement, standard diamagnetic sample (Indium) is mounted followed by switching the superconducting magnet into 'reset' mode, which locally warms the superconducting electromagnet above critical temperature and as a consequence, the effective trapped magnetic field can be nullified to a value $\leq$ 0.0001 T. The magnetic moment of indium is measured in the presence of $\mu_0 H$= 0.0002 T at 10 K and the sign and magnitude of magnetic moment are used to ensure that the trapped magnetic field is positive. All the $\textit{M}$(T) measurements are performed with 1 K/min sweep rate. Temperature dependent neutron diffraction data were collected from two axis diffractometer at D4 (Disordered materials diffractometer) in ILL, Grenoble using the wavelength of 0.4994 $\AA$ obtained by reflection of a Cu(220) monochromator \cite{chatterji2017}. Although, the idea of utilizing `hot neutrons' is remarkable for Gd containing single crystals \cite{chatterji2018polarized,bates1985magnetic, will1964neutron}, but so far any report based on powdered sample is not available in literature. The high counting rate and low  background of the D4 instrument\cite{fischer2002d4c} have enabled us to unambiguously determine the thermal evolution of magnetic structure. After calibration of the sample, the neutron diffraction intensity was normalized using a standard vanadium sample and corrected for background attenuation, multiple scattering and inelasticity (Placzek) effects. For refinement of the crystal and magnetic structure, FullProf software package was used and BasIreps\cite{rodriguez1993recent} was used for generating the irreducible representations (IR).  As reported by Lynn and Seeger\cite{lynn1990resonance},  we used a value of 9.5 fm for the coherent neutron scattering length of Gd at 0.4994 $\AA$. The absorption correction of 0.6142 was used during the
 \begin{table}[hbt]
 	\centering
 	\begin{tabular}{ |p{1.5cm}| p{2.25cm}|  p{2.25cm}|  p{2.25cm}|}
 		\hline
 		& PXRD (300 K)  & NPD  (300 K) & NPD  (3 K) \\ 
 		\hline\hline 
 		
 		Lattice parameters ($\AA$) & a = 5.3170(2) , b = 5.5204 (2) , c = 7.6084 (3)& a = 5.3152(4) , b = 5.5204 (5) , c = 7.6068 (7)& a = 5.3157(9) ,  b = 5.5147 (11) , c = 7.6009 (8)  \\\\
 		
 		Fractional co-ordinates&    &    &\\ 
 		Gd($\it{4c}$)  &      &     & \\
 		x &   -0.00579(14)    & -0.00698(51)    & -0.00718(56)  \\
 		y &   0.05720(21)    & 0.05872(28)  & 0.06021(42) \\
 		z  &  0.25000    & 0.25000 & 0.25000  \\
 		Cr($\it{4b}$) &   &    &     \\ 
 		x  &   0.50000    & 0.50000   & 0.50000 \\
 		y &  0.00000    & 0.00000  & 0.00000    \\ 
 		z &   0.00000    & 0.00000    & 0.00000    \\ 
 		O1($\it{4c}$) &   &  & \\
 		x  & 0.09141(18)  & 0.09393(29) & 0.09266(71)\\
 		y&  0.46475(23)  &  0.4043(38) &  0.40355(39) \\
 		z&  0.25000  &  0.25000 &  0.25000\\
 		O2($\it{8d}$)   &   &   &  \\  
 		x &  -0.29334(41)  &  -0.30039(22) &  -0.29892(37) \\
 		y &  0.30384(29)  &  0.29831(68) &  0.29451(79)\\  
 		z&  0.05749(18)  &  0.05608(35) &  0.05167 (29) \\  \\
 		Isotropic thermal factors ($\AA$$^2$)&     &  & \\
 		$\it{B}_{Cr}$  &  0.028(3)    & 0.062(4)  & 0.051(11)\\
 		$\it{B}_{Gd}$ &  0.031(18)     & 0.016(5)& 0.009(6)\\ \\
 		Statistical parameters &  R$_{p}$ = 20.75,  R$_{wp}$ = 15.10, R$_{exp}$ = 12.38,  $\chi$$^2$ = 1.48, Bragg R-factor 8.4242 , RF factor = 10.7931  &  R$_{p}$ = 3.31,   R$_{wp}$ = 3.64, R$_{exp}$ = 1.71 , $\chi$$^2$ = 4.52, Bragg R-factor = 0.4743, RF factor = 0.3064  &  R$_{p}$ = 0.826,  R$_{wp}$ = 1.01, R$_{exp}$ = 0.54 ,  $\chi$$^2$ = 3.46, Bragg R-factor =0.359 , RF factor = 0.189 \\
 		\hline 
 		
 		\hline \hline
 	\end{tabular}
 	\caption{Structure parameters and reliability indicators obtained from Rietveld refinement of powder XRD at 300 K and neutron powder diffraction data at 300 K and 3 K. Site occupancy is not considered as a variable during refinement process.}
 	\label{refinedparameters}
 	
 \end{table}
magnetic structure refinement procedure. For magnetic pair distribution function (mPDF) calculations, the incident $\lambda$ is 0.4994 $\AA$  which provides the maximum possible momentum transfer Q$_{max}$ = 24.3 $\AA^{-1}$. 

The room temperature crystal structure is determined by NPD and PXRD patterns presented in SM\cite{supplementarymaterial}. The experimental data is refined with help of calculated pattern generated by $\textit{Pbnm}$ space group (D$_{2h}$$^{16}$, No. 62) and lattice parameters $ \sqrt{2}a_p, \sqrt{2}a_p, 2a_p $, where a$_p$ is corresponding pseudo-cubic lattice parameter. The average values of a$_p$ are 3.8221(5) and 3.8246(2) $\AA$ obtained from refinement of NPD and PXRD data respectively. The refined values of unit cell lengths, atomic positions, thermal coefficients and reliability indicator factors are listed in Table.\ref{refinedparameters}. The good agreement between NPD and PXRD refined parameters indicates the reliability of neutron scattering  measurements. 
 
\begin{figure}[t!]
	\includegraphics[width=0.51\textwidth]{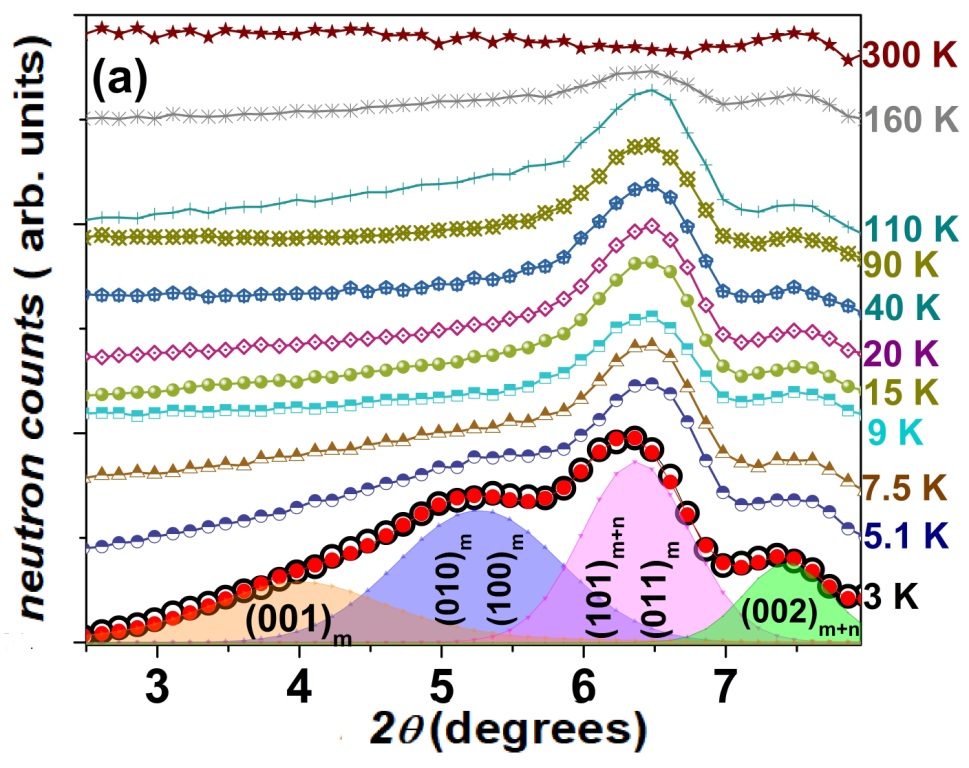}
	\includegraphics[width=0.5\textwidth]{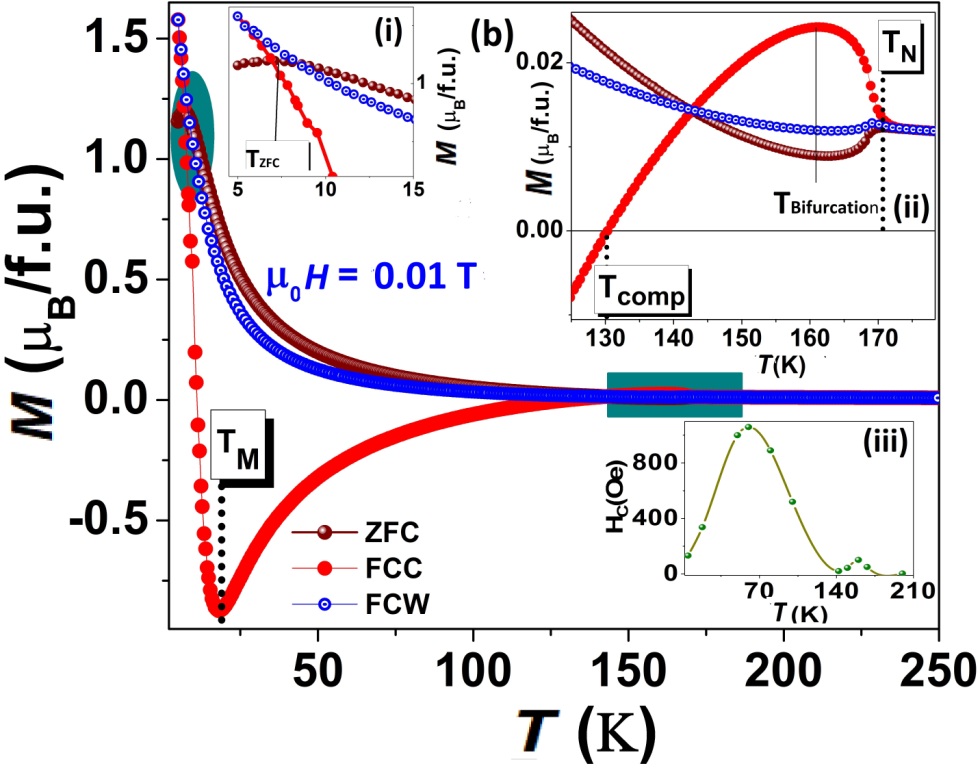}
	\caption{(color online) (a) Thermal evolution of NPD patterns. (b) $M$(T) curves in ZFC, FCC and FCW mode with measuring field $\mu_0 H$ = 0.01 T. ( Insets: (i) and (ii) show the variation of magnetization curves across $T_{ZFC}$ and $T_{comp}$ in enlarged version.(iii) Non-monotonic variation of coercivity variation with respect to temperature indicating towards the temperature dependent evolution of magnetic phases.)}
	\label{NPD_MT}
\end{figure}

Fig.\ref{NPD_MT}(a) shows the thermal evolution of NPD patterns. Below 160 K, magnetic intensity emerges at the 2$\theta$ values corresponding to (101)$_{m+n}$ and (011)$_{m}$ Bragg's reflections. Below 20 K, the slight appearance of (010)$_{m}$ + (100)$_{m}$ doublet and (001)$_{m}$ can be observed which becomes clearly visible below 7 K. It should be noted that (011), (010)/(100) and (001) reflections are prohibited in $\textit{Pbnm}$ space group and hence the intensity corresponding to these Bragg's planes appears only due to magnetic scattering. 

Temperature dependent magnetization curves measured in ZFC, FCC and FCW protocols in the presence of $\mu$$_0 H$ =  0.01 T are shown in Fig.\ref{NPD_MT}(b). The interplay of various exchange interactions between the three magnetic pairs Cr$^{3+}$-Cr$^{3+}$, Cr$^{3+}$-Gd$^{3+}$ and Gd$^{3+}$-Gd$^{3+}$ leads to a number of observed temperature driven magnetic orderings, which are nomenclatured as follows: (i) $T_N$ = 171 K; attributed to the ordering of chromium sub-lattices
 in canted antiferromagnetic structure, (ii) $T_{bifurcation}$ = 160 K; assigned with the change of the sign in slope of the three magnetization curves, FCC moments start to drop whereas, FCW and ZFC moments tend to increase (iii) $T_{comp}$ = 130 K; at which the net magnetic moment becomes fully compensated in FCC mode (iv) $T_M$ = 20 K; attributed to the sharp change in moment values and, (v) $T_{ZFC}$ = 7 K assigned to the change in slope of magnetization curve in ZFC cycle only. The temperature dependent magnetic ordering process is described by Cooke $\textit{et al.}$\cite{cooke1974magnetic} on the basis of multiple exchange interactions. The Cr$^{3+}$ ions order in canted antiferromagnetic configuration below $T_{N}$, which is attributed to Dzyaloshinskii-Moriya exchange interactions \cite{dzyaloshinsky1958thermodynamic, moriya1960anisotropic}. The canted configuration of chromium magnetic moments induces an internal magnetic field ($H_e$) at each Gd site. The total magnetic moment is assumed to be the superposition of uncompensated moments at Chromium sites ($M_{Cr}$) and paramagnetic Gadolinium magnetic moment ($M_{Gd}$) polarized due to $H_e$ in presence of applied magnetic field $H_a$, given as: 
 \begin{equation}
 \begin{split}
    M = M_{Cr} + M_{Gd} = M_{Cr} + C( H_a + H_e) /(T- \theta) 
 \end{split}
 \end{equation}
 \begin{table*}[t!]
 	\centering
 	\begin{tabular}{p{1.5cm}|c c c | c c | c c c c c c c c|} 
 		
 		\hline
 		IR &  \multicolumn{3}{c}{ symmetry elements } & \multicolumn{2}{c}{Spin modes} & \multicolumn{8}{c}{Magnetic moments at atomic positions}  \\ 
 		\hline\hline 
 		
 		&  $\tilde 2_x$& $\tilde 2_y$ & $\bar{1}$  & Cr(4b) &Gd(4c)  &  Cr$_1$&  Cr$_2$ & Cr$_3$ & Cr$_4$ & Gd$_1$ & Gd$_2$ & Gd$_3$ & Gd$_4$  \\
 		
 	 $\Gamma_1$/IR(1) &  + &  + &  + & A$_x$, G$_y$ ,C$_z$ & ., . , C$_z$ &  (u,v,w) & (-u,-v,w) & (-u,v,-w) & (u,-v,-w) & (0, 0, n ) & (0, 0, n ) & (0, 0,- n ) & (0,0,- n)  \\
 		
 		$\Gamma_2$/IR(5)&  + & -  & + & F$_x$, C$_y$  G$_z$ &  F$_x$, C$_y$, .&  (u,v,w) & (u,v,-w) & (u,-v,w)  & (u,-v,-w) & (l, m, 0 ) & (l, m ,0 ) & ( l,-m ,0 ) & (l, -m ,0) \\
 		
 		$\Gamma_3$/IR(7) &  - &  + &  + & C$_x$, F$_y$  A$_z$ & C$_x$, F$_y$, .&  (u,v,w) & (u,v,-w) & (-u,v,-w) & (-u,v,w) & (l, m, 0 ) & (l, m ,0 )  & (-l,m,0 ) & (-l,m,0) \\ 
 		
 			$\Gamma_4$/IR(3) &  - &  - &  + & G$_x$, A$_y$ F$_z$ & ., , F$_z$&  (u,v,w) & (-u,-v,w) & (u,-v,w) & (-u,v,w) & (0, 0, n ) & (0, 0, n ) & (0, 0, n ) & (0,0, n) \\
 		
 		$\Gamma_5$/IR(2) &  + &  + &  -  &   &G$_x$, A$_y$, .  &  &  & & & (l, m, 0 ) & (-l, -m ,0 ) & ( l,-m ,0) & (-l, m ,0)  \\
 		
 		$\Gamma_6$/IR(4) &  - & - &  -  &  &A$_x$, G$_y$, .  &  & &  &  &  (l, m, 0 ) & (-l, -m ,0 ) & ( -l,m ,0) & (l, -m ,0) \\
 		
 		$\Gamma_7$/IR(6) & + &  - & - &  & ., , A$_z$&   & & & & (0, 0,n ) & (0, 0,-n ) & (0, 0, -n ) & (0,0, n)  \\
 		
 		$\Gamma_8$/IR(8)  & - &  + &  -  & &  ., ., G$_z$  & & &  &  &  (0, 0,n ) & (0, 0,-n ) & (0, 0, n ) & (0,0, -n)  \\
 		
 		\hline \hline
 	\end{tabular}
 	\caption{Character table generated for $\textit{Pbnm}$ space group with $\bf{k}$ =$\bf{0}$ }
 	\label{charactertable}	
 \end{table*}	 

 \begin{figure*}[t!]
 	\includegraphics[width=0.5\textwidth]{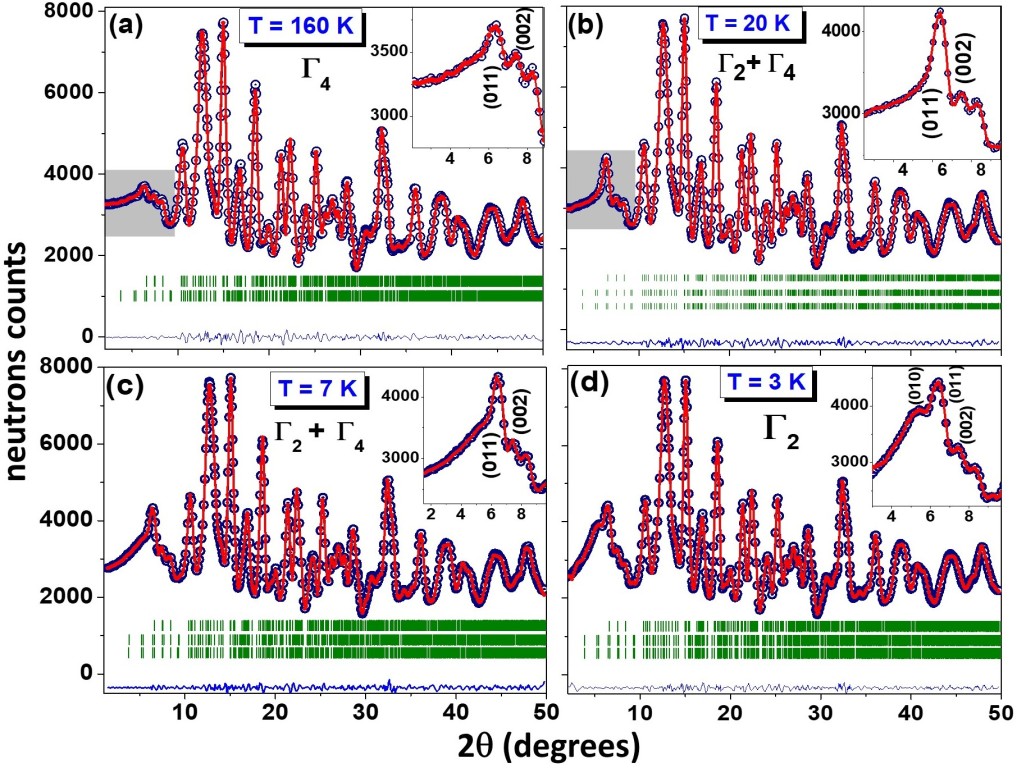}\includegraphics[width=0.53\textwidth]{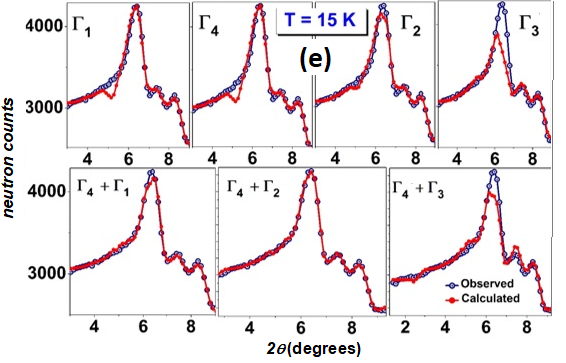}
 	
 	\caption{(a), (b), (c) and (d) : Neutron diffraction patterns along with Rietveld generated calculated patterns at temperature values $T$ = 160 K, $T$ = 20 K, $T$ = 7 K and $T$ = 3 K, respectively. Hollow circles and filled circles represent experimental and observed data points respectively. Vertical bars denote the Bragg's plane positions. Solid lines are guide to eyes. (e) Rietveld-generated patterns based on all possible magnetic configurations along with experimental pattern at $T$ = 15 K}
 	\label{refinement}
 \end{figure*}

 Using the magnetometric results, $H_e$ was estimated to be -0.55 T, where the negative sign indicates the antiferromagnetic coupling of Cr$^{3+}$ and Gd$^{3+}$ sublattices\cite{cooke1974magnetic}. K. Yoshii \cite{yoshii2001magnetic} recognized the consequence of the anti-parallel alignment of Cr$^{3+}$ and Gd$^{3+}$ magnetic moments in the observance of NM realized in FCC mode. The observed NM was remarkably stable with respect to time span as the magnetization measured at 30 K after field cooling in presence of 0.01 T applied field revealed a variation of $\sim 0.5 \%$ only on measuring after two days. The FCC magnetization was fitted using equation (1) with a value of $H_e$ = -0.15 T. 
 
 Interestingly, as shown in Fig.\ref{NPD_MT}(a), NM is observed only in FCC cycle indicating towards dependence of magnetization on the path or history in a particular measurement. To account the distinct susceptibility curves observed in FCC and FCW cycles, K. Yoshii\cite{yoshii2012magnetization} assumed the different values of $H_e$, opposite in sign but only slightly different in magnitude to empirically match the observed magnetic susceptibility. This model undoubtedly provides a significant match with the experimental data, but from the fundamental point of view, it is difficult to understand how the internal magnetic field switches its sign by changing the measuring path only. Moreover, H. J. Zhao $\textit{et al.}$, and L. Bellaiche $\textit{et al.}$ \cite{zhao2016origin, bellaiche2012simple} revealed that the effective magnetic field at rare-earth sites in $RBO_3$ materials ($B$ = Cr, Fe) is governed by the microscopic coupling of AFM ordering of B site magnetic moment and oxygen octahedral tilting. The induced magnetic moment at R site can be parallel or antiparallel with respect to the B site sub-lattice depending on the coupling constants which have characteristic value for a particular material. Therefore, it is difficult to understand why the sign of $H_e$ of chromium sub-lattice on Gd site depends on the measuring route and consequently giving rise to opposite alignment of Gd$^{3+}$ ions with respect to Cr$^{3+}$ ions. 
\begin{figure*}[t!]
	\includegraphics[width=0.5\textwidth]{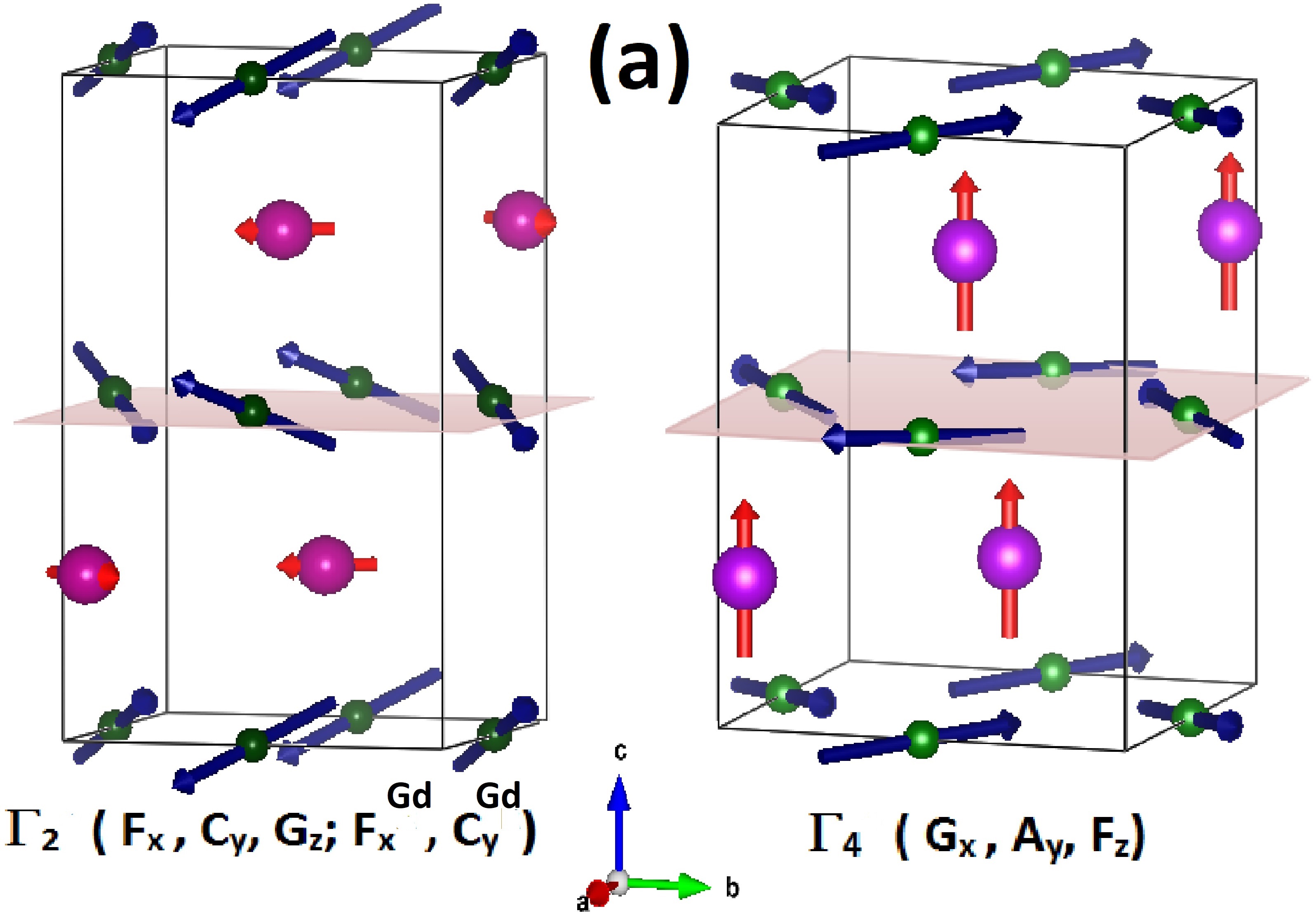}\includegraphics[width=0.48\textwidth]{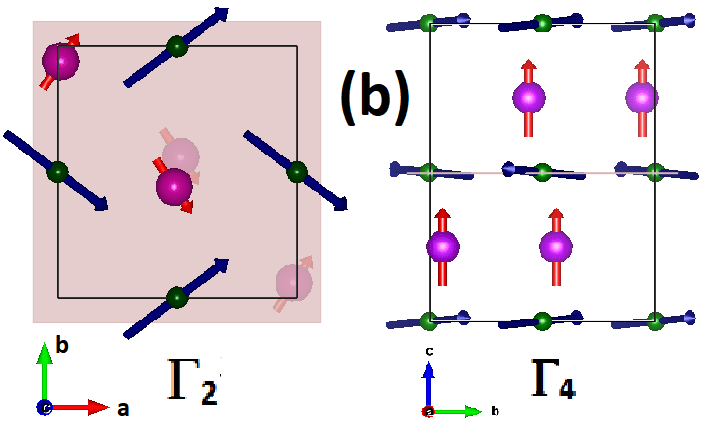}
	\includegraphics[width=0.54\textwidth]{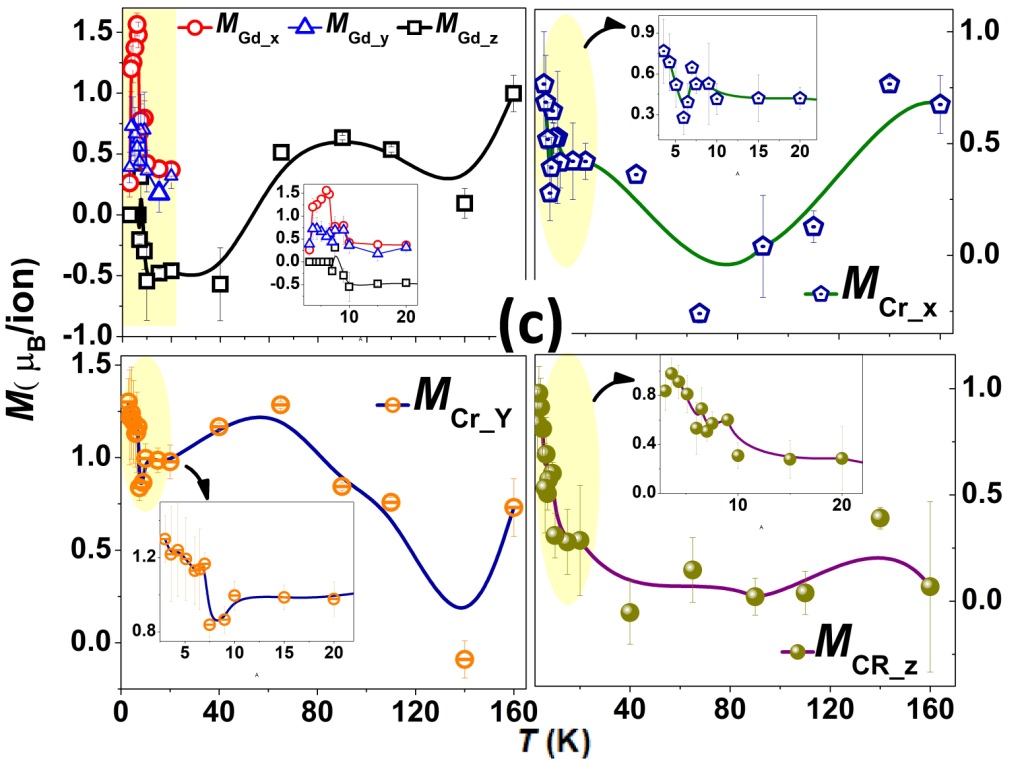}\includegraphics[width=0.5\textwidth]{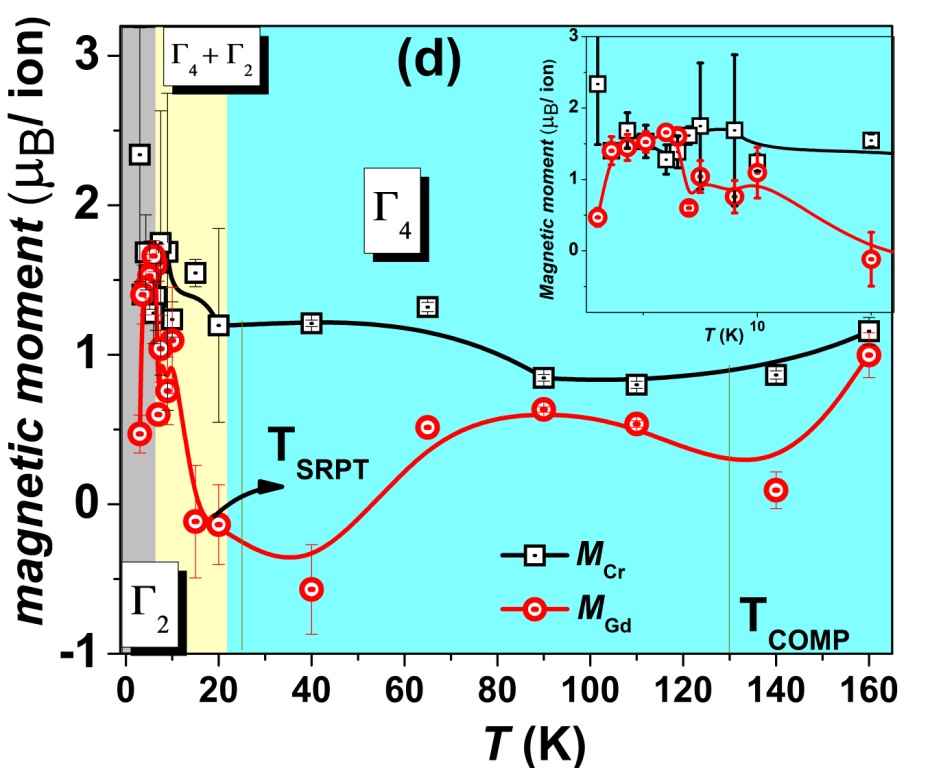}
	\caption{(a) and (b) Illustrations of $\Gamma_4 \equiv G_x, A_y, F_z$ and $\Gamma_2 \equiv F_x, C_y, G_z$ magnetic spin configurations (c) Variation of magnetic moment components along x,y, and z directions. (d) Temperature driven phase diagram of GdCrO$_3$ along with variation of total Cr and Gd ionic moments with respect to temperature.}
\end{figure*}\\
 \begin{figure}[h!]
	\includegraphics[width=0.70\textwidth]{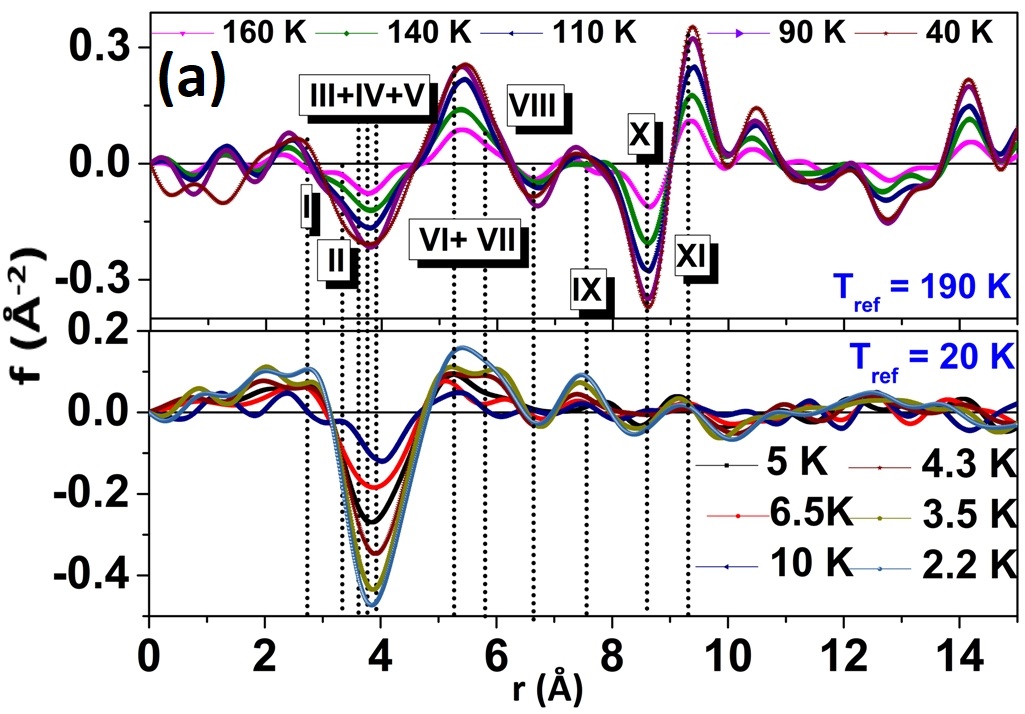}\includegraphics[width=0.32\textwidth]{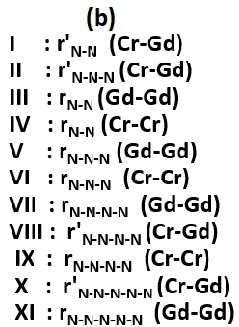}
	\includegraphics[width=0.47\textwidth]{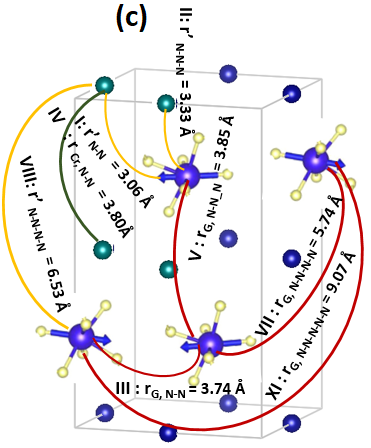}
	\caption{\label{figmPDF}(color online)(a) mPDF calculations with reference of two different base temperatures T = 190 K and 20 K.(b) List of various bond lengths, where $r_{Cr}$, $r_{Gd}$ and $r'$represent the Cr-Cr, Gd-Gd and Cr-Gd bond lengths, respectively. (c)Illustration of various atomic distances in single unit cell of GdCrO$_3$. Blue (large) and green (small) spheres represent Gd and Cr atoms respectively. For the sake of clarity inter-atomic distances corresponding to VI: $r_{N-N-N}$(Cr-Cr), IX: $r_{N-N-N-N-N}$ (Cr-Cr) and X: $r'$$_{N-N-N-N-N }$ are omitted.}
\end{figure}    
	In RCrO$_3$ family, the crystallographic and magnetic unit cells are identical, i.e., magnetic structure can be generated by $\textbf{k = 0}$ propagation vector. Chromium atoms are at 4$\textit{b}$ Wyckoff positions where the atomic co-ordinates of these atoms are $(\frac{1}{2}, 0, 0 )$; $(\frac{1}{2}, 0, \frac{1}{2} )$; $(0, \frac{1}{2}, \frac{1}{2})$; $(0, \frac{1}{2}, 0 )$. Gadolinium atoms are at 4$\textit{c}$ sites with atomic positions given as: $(x, y, \frac{1}{4})$; $(-x, -y, \frac{3}{4})$; $( \frac{1}{2} +x, \frac{1}{2} - y, \frac{3}{4} )$;$( \frac{1}{2} - x, \frac{1}{2} + y,  \frac{1}{4} )$. For $\textit{Pbnm}$ space group, the independent symmetry elements are two two-fold screw axes $\tilde 2_x$ and $\tilde 2_y$ at $(x, \frac{1}{4}, 0 )$ and $( \frac{1}{4}, y, \frac{1}{4} )$, respectively and the inversion centre $\bar{1}$ at the point (0,0,0). It should be noted that $\tilde 2_z = \tilde 2_x . \tilde 2_y$ and henceforth, is not considered as an independent symmetry element. For the sake of convenience, the linear combinations of spin vectors S$_j$ (j = 1-4) which transform into themselves under the operation of symmetry elements defined as:
	\begin{center}
	$F = S_1 + S_2+ S_3 + S_4$ \\
	$A = S_1 - S_2 - S_3 + S_4$ \\
	 $ C = S_1 + S_2 - S_3 - S_4 $\\
	 $ G= S_1 - S_2 + S_3 - S_4 $ 
	\end{center}
form the basis vectors \cite{bertaut1968representation,bertaut1962lattice}. The definite transformation properties are called a representation. The representation analysis studied by Bertaut \cite{bertaut1968representation} helps to assign the irreducible representations (IR) of the space group to a known magnetic structure. The four allowed combinations for chromium 4$\textit{b}$ sites are denoted as $\Gamma_i$ (i = 1,2,3 and 4) along with the BasIreps generated IRs and corresponding basis vectors are listed in Table.\ref{charactertable}. In case of Gd (4$\textit{c}$) sites, there are total 8 irreducible representations, comprising of the additional four representations denoted as $\Gamma_j$ (j= 5-8). The transformation properties of these representations and value of these components (moments) are also listed in Table \ref{charactertable}. The moments of chromium and gadolinium atoms are represented as (u,v,w) and (l,m,n), respectively. The reducible representation belonging to chromium 4$\textit{b}$ sites can be written as the linear combinations of irreducible matrices $\Gamma_i$s:
$$ 3\Gamma_1  + 3\Gamma_2 +3\Gamma_3  +3\Gamma_4  $$
  
The magnetic structure is modelled with respect to the Rietveld refined patterns using FullProf. It is observed that the magnetic structure below $T_N$ can be generated with $\textbf{k=0}$ propagation vector and $\Gamma_4$ or G$_x$, A$_y$, F$_z$ configuration as shown in Fig.\ref{refinement}. The $\Gamma_4$ configuration is observed to be the most reliable magnetic structure for the temperature values 20 K $\leq T \leq T_N$. Below 20 K, no individual allowed IR is able to provide a reliable match with the experimental data, so we have generated the patterns with intermediate phases formed by possible combinations of two IR's. It may be argued that the inclusion of an additional phase and henceforth providing more degrees of freedoms should improve the quality of match anyhow. To justify the choice of proper combination, we have presented the generated patterns with all individual IR's and their possible combinations as shown in Fig.2(e). It can be seen that $\Gamma_2$ + $\Gamma_4$ $\equiv$ $\Gamma_{24}$ unambiguously provides the most reliable match with the experimental profile and hence assigned as the magnetic structure for 7 K $\le$ T $\le$ 20 K as shown in Figs.2(b) and (c).  Below 7 K, a clear 
enhancement in magnetic intensity is observed corresponding to (010)$_m$ + (100)$_m$ doublet and (001)$_m$  Bragg's plane (Fig. 2(d)). Based on the calculations of Shamir $\textit{et al}$; \cite{shamir1981magnetic} the appearance of (010)$_m$ + (100)$_m$ doublet is attributed to the ordering of Gd$^{3+}$ moments. The magnetic structure below 7 K is generated  with ordering of chromium ionic sites (4b) in $\Gamma_2$ (F$_x$, C$_y$, G$_z$) configuration along with Gd$^{3+}$ (4c) ordering in $\Gamma_2$ (F$_x$, C$_y$) configuration as shown in Fig. 2(d). The spin arrangements in both $\Gamma_2$ and $\Gamma_4$ are illustrated in Figs.3(a) and (b). In $\Gamma_4$ (G$_x$, A$_y$, F$_z$) configuration, the magnetic moments of nearest neighboring Cr sites follow G-type ordering along x$\parallel$a \\\\\\\\\\\\\\\\\\\\\\\\\\\\\\\\\\\\\\\\\\\\\\ direction, A-type along y$\parallel$b and ferromagnetic ordering along z$\parallel$c direction, resulting in an uncompensated weak moment along z direction. Similarly, in $\Gamma_2$ (F$_x$, C$_y$, G$_z$) configuration, x$\parallel$a componentsof magnetic moment order in ferromagnetic configuration henceforth an uncompensated moment along x crystallographic axis. In the temperature regime of 7 K $\leq T \leq $ 20 K, the direction of uncompensated spins reorient from z$\parallel$c axis (20 K ) to x$\parallel$a axis (7 K) forming a spin reorientation phase transition (SRPT). Now onwards, we shall denote the unidentified magnetic transitions $T_{M}$ and $T_{ZFC}$ as $T_{SRPT}$ and $T_{Gd}$ respectively. It is noteworthy that no crystal or magnetic structural modification is observed across $T_{bifurcation}$ and $T_{comp}$; and thus we infer that origin of magnetization reversal is not associated with magnetic phase transitions. 
 
 The variation of components of magnetic moment along different crystallographic directions obtained from Rietveld refinement is shown in Fig.3(c).  The component of  chromium magnetic moment along z direction is very small in $\Gamma_4$ phase and moment is confined in a-b plane only. The evolution of magnetic moments along x and y axes across $T_{bifurcation}$ and $T_{comp}$ is almost opposite to each other, indicating an in-plane rotation of magnetic moment from a$\parallel$x to b$\parallel$y axis. All  the three moments reveal a significant drop in absolute moment value in the vicinity of $T_{Gd}$ and the moment values again start to enhance when the system is completely transformed into $\Gamma_2$ phase below $T_{Gd}$, as shown in respective insets in Fig. 3(c). The one dimensional temperature driven phase diagram along with variation of total moment at chromium and gadolinium sites are shown in Fig. 3(d). The net magnetic moment of Gd$^{3+}$ ionic sites is although sensitive to the temperatures $T_{comp}$ and $T_{bifurcation}$, it is negative only for $T_{Gd}$ $\leq$ $T \leq$ 40 K and not for the entire negative moment regime described by the bulk magnetization results. It reveals that polarized magnetic moment of Gd does not make a significant contribution to the phenomenon of negative magnetization. 
                                                                                           
As the average long-range structure does not indicate towards the NM, we aim to look for the nano-scale short range correlations.  We compute the real space correlation functions or magnetic pair distribution functions (mPDF) by Fourier transformation of the total magnetic scattering intensity into real space as shown in Fig.4(a). The mPDF technique is sensitive to short-range correlations as the diffuse scattering is also accounted besides the Bragg's scattering \cite{frandsen2014magnetic, frandsen2015magnetic}. Typically, a particular peak position in mPDF corresponds to a pair seperation distance, slope of linear baseline depends on the spin orientation, the sign of peak attributes the nature of ordering and the peak height is the function of components aligned perpendicular to connecting axis joining them. The contribution to the mPDF due to a pair of spins S$_i$ and S$_j$ separated by a distance r$_{ij}$ is given as
$$ f_{ij} = C [ A_{ij} \delta(r-r_{ij}) /r+ B_{ij} \Theta(r_{ij}-r)r/r_{ij}^3] $$
 where C is correlated with spin quantum number, $\Theta$ is Heavside step function, $A_{ij}$ and $B_{ij}$ are correlation coefficients determined by the alignment of spins, generally positive for ferromagnetic type alignment and negative for antiferromagnetic arrangement\cite{frandsen2014magnetic}. To investigate the correlations between short scale interactions and negative magnetization, we have calculated mPDF for difference diffraction patterns with reference to two base lines, one at 190 K which is above Ne\'el temperature and second at 20 K ($T_{SRPT}$). The first two features (I and II) correspond to the exchange interactions between chromium - gadolinium 6 first and 12 second nearest neighboring pairs (NN and NNN) respectively. The third feature (III+ IV + V) arising because of the convoluted effects of Gd-Gd NN, Cr-Cr NN and Gd-Gd NNN interactions, is well defined and demonstrates resultant strong antiferromagnetic coupling. Similarly, the fourth feature (VI + VII) originating because of Cr - Cr NNN and Gd-Gd NNNN exchange interaction is also prominent but indicates towards ferromagnetic ordering. The distinct peak VIII is attributed to the Cr-Gd third nearest neighbour interaction (NNNN), which is ferromagnetic in nature but intensity is reduced as a factor of 1/r. Features IX and X correspond to Cr-Cr third next neighboring (NNNN) and Gd-Cr fourth nearest neighboring (NNNNN) interactions, respectively. Peak XI represents the Gd-Gd fourth next nearest neighboring (NNNNN) interactions. The corresponding atomic distances are illustrated in Fig.4(b). 
 
Analysing the mPDF for two different types of spins is ambiguous and hence here we will consider the correlations in disordered Gd$^{3+}$ spins only. In the NM regime, there is significant local Gd$^{3+}$-Gd$^{3+}$ 3-D antiferromagnetic interactions ranging up to the four nearest neighbors (NNNNN) distance or $\sim$ 9$\AA$, even though the long range structure of Gd is paramagnetic. The intensity of these Gd-Gd correlation peaks varies in the proportionally with the coordination number. Just below $T_{SRPT}$, when the magnetization flips to become positive in FCC cycle, the intensity corresponding to second and third nearest neighbor interaction of Gd$^{3+}$ drops. Noteworthily, the intensity attributed to Gd-Gd NNNNN interaction is now significantly reduced whereas, the feature belonging to Gd NN interaction is enhanced, suggesting that the local Gd$^{3+}$ correlations now cease. 
  \begin{figure}[b!]
	\includegraphics[width=0.4\textwidth]{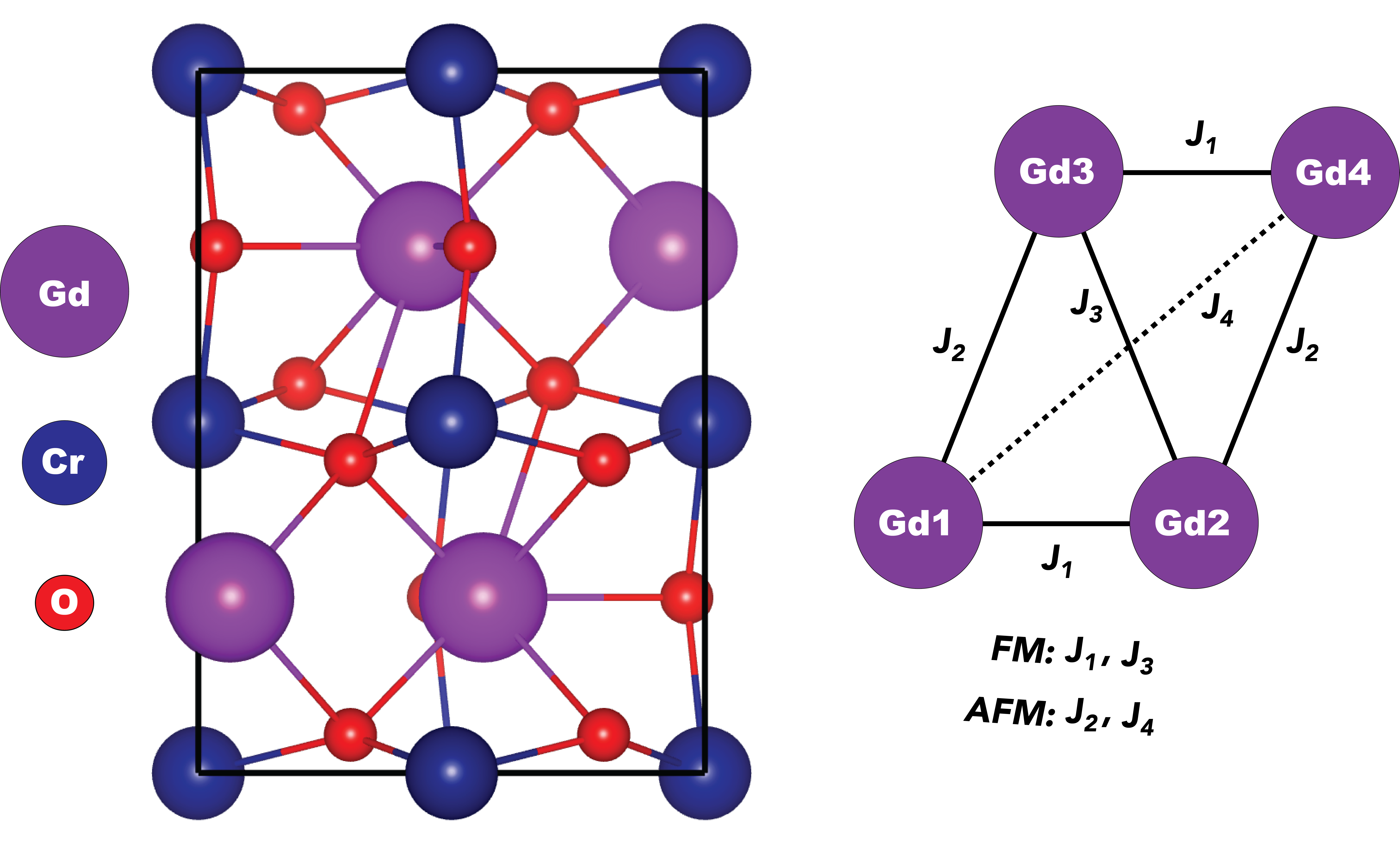} 	
	\caption{\label{figModel} (Left) Unit cell of \GCO and (Right) pathways of magnetic exchange between Gd ions. $J_i$'s are the magnetic exchange strengths. }
\end{figure}

To elucidate the nature of short range correlations of Gd$^{3+}$ ions, we have modelled the local interactions of Gd ions within the framework of isotropic Heisenberg Hamiltonian. The model that we consider here involves Gd sublattice in presence of an effective magnetic field $H_e^{z}$, which is the sum of externally applied field and the internal field due to the ordered Cr sublattice. The spin Hamiltonian to describe the system is given by,

 \begin{eqnarray}
\label{eqHeisH}
  \hat{H}=-\sum_{i<j} J_{ij} \hat{\vec{s}}_i \cdot \hat{\vec{s}}_j - g\mu_BH_e^{z}\sum_i \hat{s}_i^z
 \end{eqnarray}

 \noindent where, the first and second terms in the equation correspond to Heissenberg exchange and Zeeman terms respectively. In eqn. \ref{eqHeisH}, $J_{ij}$ is the magnetic exchange between sites $i$ and $j$, $\hat{\vec{s}}_i$'s are the site spin operators, $g$ is the gyromagnetic ratio taken to be 2.0 and $\mu_B$ is the Bohr magneton. Positive and negative values of $J_{ij}$ correspond to ferromagnetic and antiferromagnetic interactions respectively. As one can see from the Zeeman term, a positive value of magnetic field will stabilize the spin states with positive total $M_s$ values. In presence of negative effective magnetic field, spin states corresponding to negative magnetization are stabilized relative to those with positive magnetization, thus resulting in overall negative magnetization. 

The magnetic exchange pathways within the unitcell of the sytem is shown in the figure \ref{figModel}. From the unitcell topology, six Gd-Gd magnetic exchange pathways can be identified. Correspondingly, the model Hamiltonian can be written as, 

 \begin{eqnarray}
\label{eqHeisHmoda}
  \hat{H}&=&-J_{12} \hat{\vec{s}}_1 \cdot \hat{\vec{s}}_2 -J_{13} \hat{\vec{s}}_1 \cdot \hat{\vec{s}}_3 -J_{14} \hat{\vec{s}}_1 \cdot \hat{\vec{s}}_4 \nonumber \\
&&-J_{23} \hat{\vec{s}}_2 \cdot \hat{\vec{s}}_3 -J_{24} \hat{\vec{s}}_2 \cdot \hat{\vec{s}}_4 -J_{34} \hat{\vec{s}}_3 \cdot \hat{\vec{s}}_4 \\
\label{eqHeisHmodb}
&=&-J_{1} \hat{\vec{s}}_1 \cdot \hat{\vec{s}}_2 -J_{2} \hat{\vec{s}}_1 \cdot \hat{\vec{s}}_3 -J_{4} \hat{\vec{s}}_1 \cdot \hat{\vec{s}}_4 \nonumber \\
&&-J_{3} \hat{\vec{s}}_2 \cdot \hat{\vec{s}}_3 -J_{2} \hat{\vec{s}}_2 \cdot \hat{\vec{s}}_4 -J_{1} \hat{\vec{s}}_3 \cdot \hat{\vec{s}}_4
 \end{eqnarray}
\begin{table}[t!]
	\caption{\label{tabBondInfo}Gd-Gd distances, Gd-O-Gd bond angles and the corresponding exchange type in the unitcell of \GCO.}
	\centering
	\begin{tabular}{|c|c|c|c|c|} \hline
		Ion pair & Distance ($\AA$) & Bond angle & Exchange & Interaction\\\hline 
		Gd1-Gd2 & 3.741(5) & 84.13 (7)$^o$ & $J_1$ & FM \\
		Gd1-Gd3 & 5.741 (2) & 177.30 (3) $^o$& $J_2$ &  AFM \\
		Gd2-Gd3 & 3.861 (6) & 79.55 (19) $^o$ &  $J_3$ & FM \\
		Gd1-Gd4 & 8.905 (5) & --- &  $J_4$ &  AFM \\\hline
	\end{tabular}
\end{table}	
The sign and strength of various magnetic interactions in the unit cell can be ascertained from the Gd-O-Gd bond lengths ($d$) and angles ($\theta$) respectively, and the same are presented in the table \ref{tabBondInfo}. The bond information suggests that there are only four unique magnetic exchange constants, $J_1$ through $J_4$ as shown in eqn. \ref{eqHeisHmodb}. Among these, the bond angles corresponding to $J_1$ and $J_3$ are approximately close to $90\degree$ and are expected to be ferromagnetic (positive values of $J$). Whereas, the bond angles for $J_2$ is closer to $180\degree$, suggesting an antiferromagnetic interaction for this pathway. The exchange constant $J_4$ is taken to be weakly antiferromagnetic. A 1/$d$ dependence can be assumed for the strength of magnetic interactions and the magnitude of strongest interaction is taken to be 1. Correspondingly, the starting values of exchange constants are fixed to $J_1$=1.0,  $J_2$=-0.75, $J_3$=0.95 and $J_4$=-0.47. In the above equation, the Zeeman term is not included for the sake of simplicity. This term is diagonal as it involves only the $z$-component of spin operator, which will shift the energy eigenvalues by $-g\mu_BH_e^{z}M_S$, where $M_S$ is the total magnetization of the eigenstate. This term is added separately to the energy eigenvalues while computing the magnetization. 
 \begin{figure}[hbt]
	\includegraphics[width=0.46\textwidth]{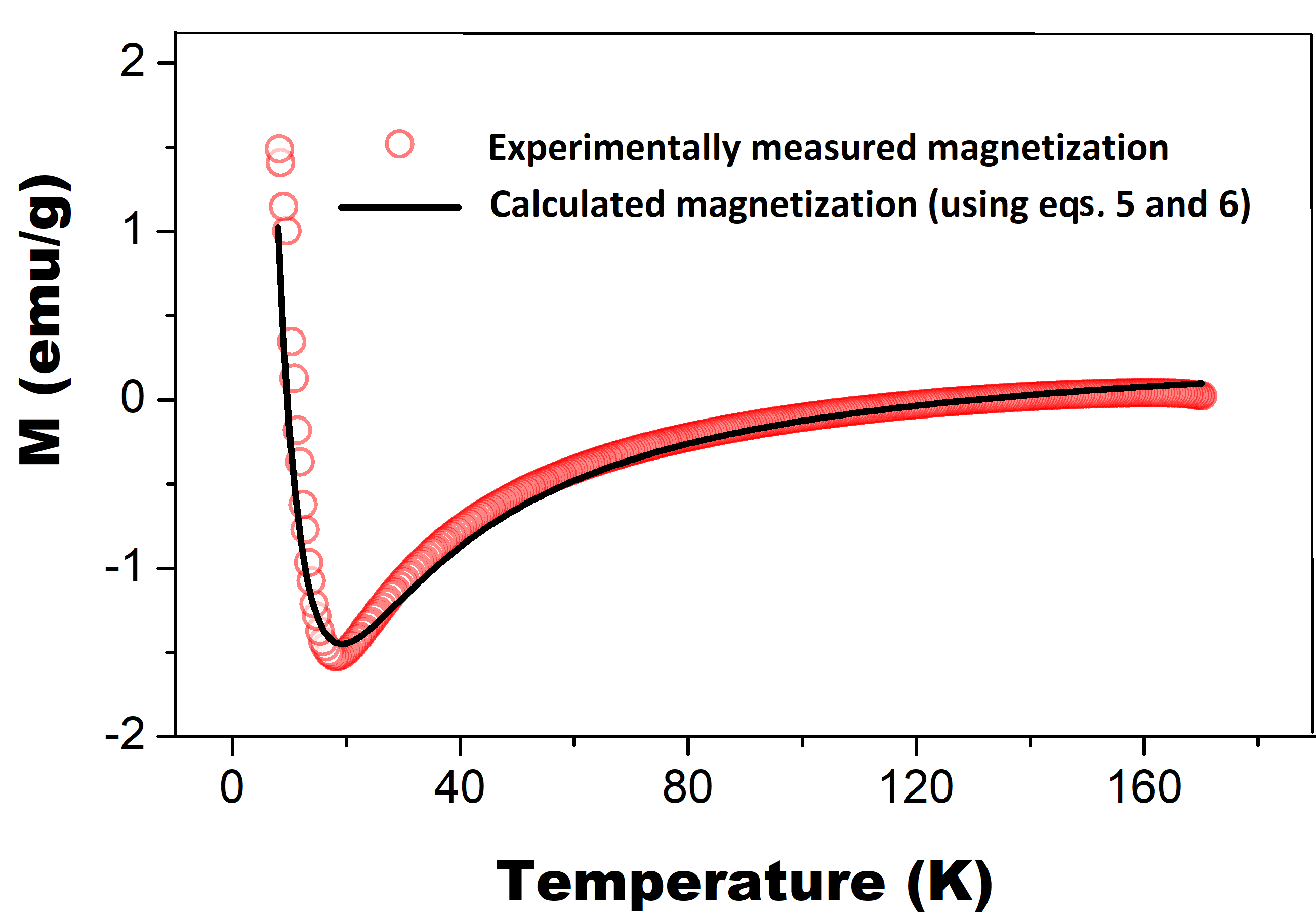} 	
	\caption{ Experimental magnetization data of \GCO and theoretical fit for $J_1$=2.3 K; $J_2$=-1.66 K; $J_3$=2.19 K and $J_4$=-0.23 K; $H_e^{z}$=-0.18 T and $D$=0.18 K.}
	\label{figFit} 
\end{figure}
\begin{figure*}[hbt]
	\includegraphics[width=\textwidth]{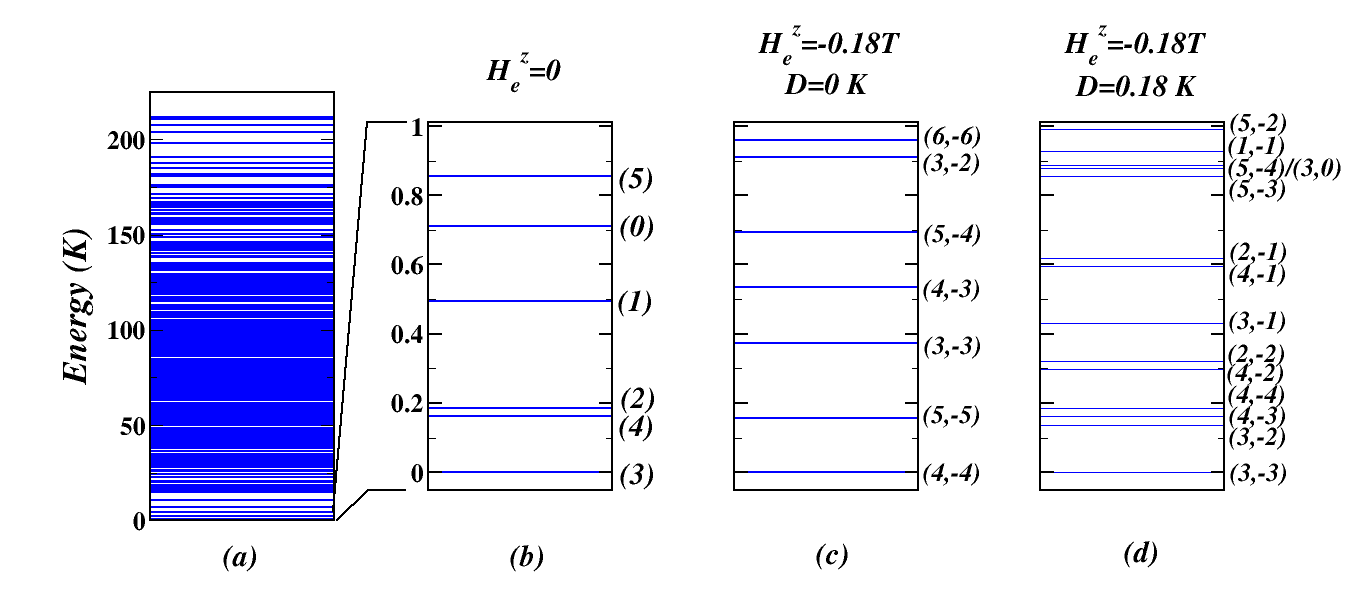} 	
	\caption{\label{figEspec}  (a) Calculated eigenspectrum of the model in equation.\ref{eqHeisHmodb} for $J_1$=2.3 K; $J_2$=-1.66 K; $J_3$=2.19 K and $J_4$=-0.23 K. Zoom of eigenspectrum in the energy range 0 to 1 K  for (b) $H_e^{z}$=0, (c) $H_e^{z}$= -0.18 T and (d) $H_e^{z}$= -0.18 T, $D$=0.18 K. In the $H_e^{z}$=0 case, the spin states are $2S+1$ degenerate and hence are indexed only by their total spin ($S^{tot}$). For the $H_e^{z}$=-0.18 T cases, the degeneracy is lifted and the states are indexed with ($S^{tot},M_S^{tot}$).}
\end{figure*} 
The model Hamiltonian presented in eqn. \ref{eqHeisHmodb} can be constructed in the basis of total spin   ($S$) or total $z$-component of the total spin ($M_S$), as the corresponding operators $\hat{S}^2$ and $\hat{S}^z$ commute with $\hat{H}$. In the present case, the $H$ matrix is constructed in the constant $M_S$ basis.  Numerically solving the model Hamiltonian to obtain all the spin eigenstates ($E(S,M_S)$) is discussed elsewhere\cite{raghunathan2008theoretical}. Since, the model Hamiltonian commutes with $\hat{S}^2$ and $\hat{S}^z$ operators, the eigenstates of the Hamiltonian in eqn.  \ref{eqHeisHmodb} are also simultaneously the eigenstates of these operators. Hence, the total spin ($S$) and the total $M_S$ of every eigenstate are obtained from the expectation values of the $\hat{S}^2$ and $\hat{S}^z$ operators. The canonical partition function is used to compute the magnetization arising from the Gd sublattice as a function of temperature (T) for a given value of effective magnetic field, $H_e^{z}$ (eqn. \ref{eqnMag}).

 \begin{eqnarray}
\label{eqnMag}
M_{Gd}(T)=N_A g \mu_B \frac{\sum_S\sum_{M_S}M_Se^{-\frac{E_t(S,M_S)}{k_BT}}}{\sum_S\sum_{M_S}e^{-\frac{E_t(S,M_S)}{k_BT}}}
 \end{eqnarray}

In the above equation, the total energy of the eigenstate is given by, $ E_t(S,M_S)=E(S,M_S)-g\mu_BH_e^{z}M_S+DM_S^2$ and $N_A$ is the Avagadro number. The additional terms in the energy expression are the Zeeman and anisotropy energy respectively. The total magnetization of the system is given by, 
 \begin{eqnarray}
\label{eqntotMag}
 M(T) = M_{Cr}(T) + M_{Gd}(T) + M_{P}(T)+M_{D}
 \end{eqnarray}
\noindent where, the terms on the right hand side (RHS) of the equation are the magnetization arising from Cr sublattice, Gd sublattice, paramagnetic and diamagnetic impurity contributions, respectively. The third term on the RHS refers to the magnetization of the residual uncorrelated moments. This follows approximately $1/T$ dependence and becomes significant at very low temperatures. It should be noted that due to very small canting angle the uncompensated magnetic moment of Cr$^{3+}$ ions is very low in $\Gamma_4$ configuration\cite{tripathi2017evolution}. The estimated values of $M_{Cr}(T)$ ranges from 1.34(2)-9.22(7) $\times$ 10$^{-3}$ emu/g which is orders of magnitude smaller than the total moment. Hence, for simplicity the magnetization contribution from the Cr sublattice can be neglected while fitting the total magnetization data. The third term in eqn.\ref{eqntotMag} corresponds to the magnetization due to free Gd spins and is of the form $CH_e^{z}/T$, where C is the Curie constant. As this term has $1/T$ dependence, the paramagnetic contribution is significant at very low temperatures. 
 
The magnetization data is fitted iteratively by changing the relative strength of $J_2$, $J_3$ and $J_4$ keeping $J_1$=1. The fitted magnetization data is shown in figure \ref{figFit}. The best fit paramaters yield $J_1$=2.3 K; $J_2$=-1.66 K; $J_3$=2.19 K and $J_4$=-0.23 K. The effective magnetic field used for the fit is  $H_e^{z}$=- 0.17 T, or the internal field due to Cr sublattice on Gd ions is -0.18 T. The anisotropy constant used for the fit is $D$=0.18 K. The ground state (GS) of the model in the absence of any magnetic field corresponds to spin septet ($S_{GS}=3$) due to the frustration induced by the exchange intraction.
\begin{figure*}[t!]
	\includegraphics[width=\textwidth]{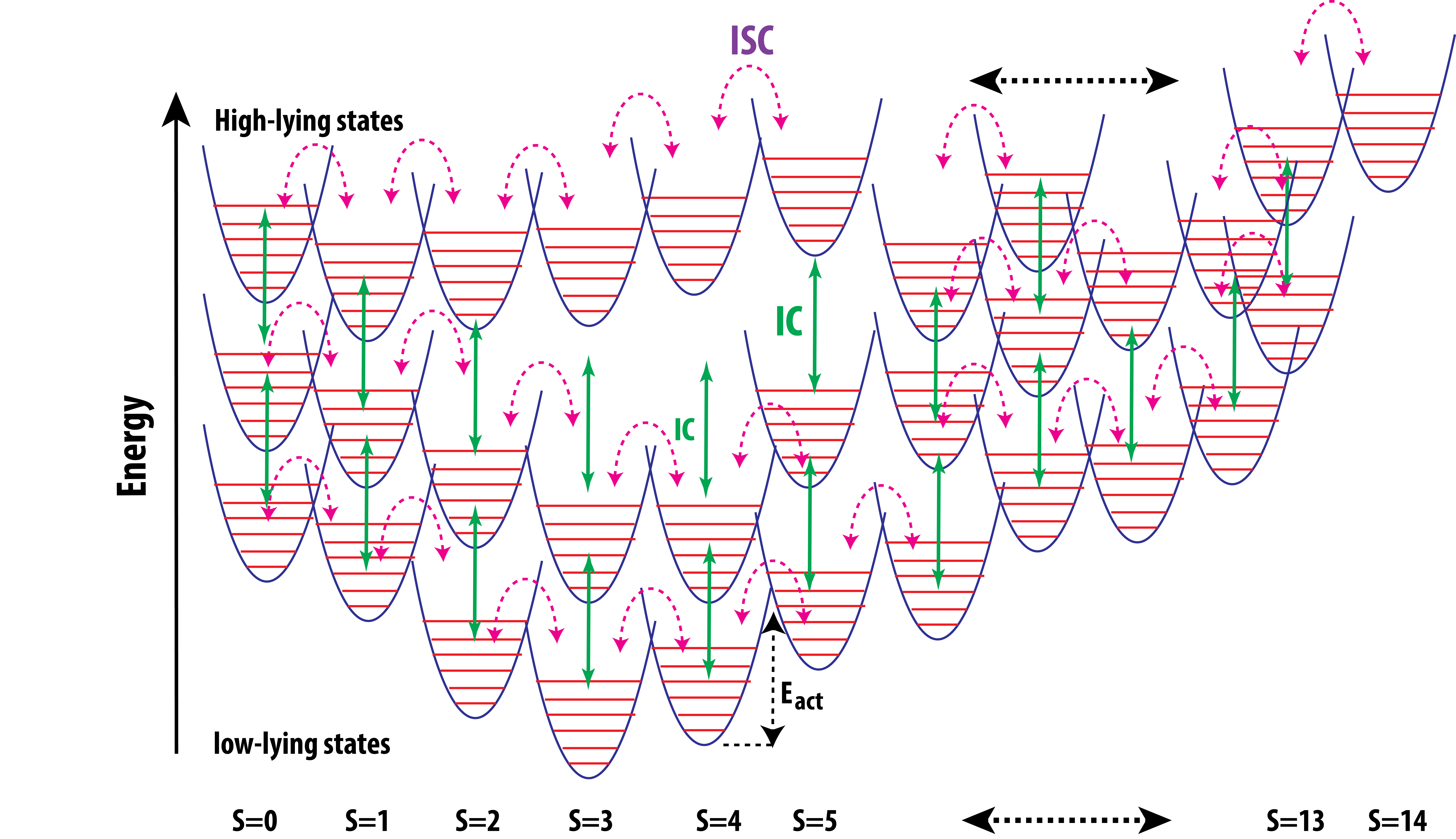} 	
	\caption{\label{figEspec} Schematic of kinetic process that involves internal conversion (IC) and intersystem crossings (ISC). The potential energy surface corresponding to various spin manifolds of the system is shown. Each spin manifold is connected to the other by an activation barrier $E_{act}$.}
\end{figure*}
The energy spectrum of the model for $H_e^{z}=0$, $H_e^{z}$= -0.18 T and $H_e^{z}$=- 0.18 T; D=0.18K is shown in the fig. \ref{figEspec}. The low-lying spectrum of the model consists of very closely spaced excited states belonging to total spin 4, 2, 1, 0 and 5 within an energy gap of 1 K from the $S=3$ GS. The excitation gaps are small due to the very weak exchange interactions present in the system. In presence of negative magnetic field $H_e^{z}$, the $2S+1$ degeneracy of the spin states is lifted. The states corresponding to the negative magnetization are stabilized relative to the positive ones and $S^{tot}=4,~M_S^{tot}=-4$ becomes the ground state. The Boltzmann weights for these negative magnetization states are large at low temperatures, which lead to negative values of total magnetization. This situation is changed in presence of the anisotropy term $D$, in which case the GS corresponds to $S^{tot}=3,~M_S^{tot}=-3$. This is because the positive nature of $D$ destabilizes the states corresponding to both positive and negative magnetizations. This destabilization is greater for larger $M_S$ values and hence the $S^{tot}=3,~M_S^{tot}=-3$ stabilizes to GS. Even in presence of anisotropy, the low-lying states of the spectrum are completely dominated by states with negative $M_S$ values. It should be noted that the paramagnetic contribution obtained from the fit is very large with a Curie constant of $C$=70 emu K/g. This signifies that at low temperatures the free ion contribution to the total magnetization outweighs that of the correlated Gd spins resulting in a crossover to positive magnetization. This large paramagnetic contribution can be corroborated with the very short range nature of magnetic correlations present in the system.

Our model thus provides insights on the complex nature and strength of magnetic exchange in \GCO system. Though the presence of negative internal field created by the Cr sublattice leads to negative magnetization during the field cooled cycle, our experimental findings show that the magnetization remains positive during the warming cycle. Here we argue the possibility that the system having different initial population distribution in closely lying energy states and its distinct evolution with temperature leads to an entirely different path during the warming cycle. When the system is warmed gradually, the spins are  excited from the GS spin manifold through a series of energy barriers. This is controlled by the internal conversion and inter-system crossings. Internal conversion refers to excitation of spins within same spin vector, whereas the inter-system crossings correspond to crossover to a different total spin, obeying the spin selection rules ($\Delta S=\pm 1$). A purely paramagnteic state is achieved when the thermal energy is large enough to overcome the activation barriers to populate the high-lying states of the energy spectrum. Assuming that at absolute zero, all the spins are ordered and are stabilized in the $S^{tot}=3,~M_S^{tot}=-3$ GS, the warming cycle can be tracked by a set of kinetic rate equations which include internal conversions and inter-system crossings to various spin states\cite{raghunathan2008kinetic}. In this case, the concentration of various spin species and hence the mangetization of the system depends on the solutions of  the rate equations as a function of temperature and time. Such a model provides an understanding of the positive magnetization during the FCW cycle.

In summary, we have constructed the temperature driven magnetic phase diagram and discussed the origin of negative magnetization (NM) in GdCrO$_3$ by utilizing the high energy $\lambda$ = 0.4994 $\AA$ neutrons to overcome the huge absorption of natural Gd ions. Unambiguously, three distinct magnetic phase transformations are observed : chromium moments ordering in G$_x$, A$_y$, F$_z$ with uncompensated moment along z$\parallel$c direction below Ne\'el temperature $T_N$ = 171 K, rotation of chromium weak ferromagnetic moments along x$\parallel$a crystallographic direction comprising  F$_x$, C$_y$, G$_z$ spin configuration and Gd moments ordering in  G$_y$, A$_y$ configuration below T$_{Gd}$ = 7 K. In the vicinity of spin reorientation phase transition (SRPT) [7 K $\leq T \leq $ 20 K], an intermediate mixed spin configuration is observed. Unexpectedly, no significant changes in long range magnetic structure is observed across $T_{compensation}$ and $T_{bifurcation}$, suggesting that the NM is not associated with long range magnetic phase transformations. Short range Gd$^{3+}$ correlation functions derived by mPDF calculations reveal the significant AFM correlations up to third nearest neighbor distance or $\sim$ 9 $\AA$, which cease below $T_{SRPT}$. Based on these observations, we have modelled the system with a model spin Hamiltonian. Results from our model calculations show that the exchange interactions between the Gd spins are extremely weak leading to set of closely spaced energy levels above the ground state. Competing exchange pathways in the system result in spin frustration leading to non-zero spin ground state. Presence of a negative effective magnetic field stabilizes the states with negative total-$M_S$ values. The path dependency in observance of NM is understood as a consequence of the distinct population distribution in various closely spaced excited states with respect to cooling and warming paths. 

 We are grateful to Dr. Juan Rodriguez-Carvajal for reading the whole manuscript and providinsg salient comments. ILL, Grenoble is thanked for providing the access to neutron diffraction facility. 

\bibliography{References}

\end{document}